\documentclass[sigconf]{acmart}
\AtBeginDocument{%
  }

\setcopyright{acmlicensed}
\copyrightyear{2025}
\acmYear{2025}
\acmDOI{XXXXXXX.XXXXXXX}
\acmConference[Conference acronym 'XX]{Make sure to enter the correct
  conference title from your rights confirmation email}{June 03--05,
  2025}{Woodstock, NY}
\acmISBN{978-1-4503-XXXX-X/2025/10}

\newcommand{\model}{DynaCausal}

\usepackage{enumitem}
\usepackage{algorithm}

\usepackage{cleveref}
\usepackage{algorithmic}
\usepackage{graphicx}
\usepackage{textcomp}
\usepackage{pifont}
\usepackage[T1]{fontenc}

\usepackage{etoolbox}
\usepackage[table,xcdraw]{xcolor} 
\usepackage{tcolorbox}
\usepackage{fancybox}
\usepackage{subcaption}
\usepackage{caption}
\usepackage{multirow}
\usepackage{multicol}
\usepackage{makecell}
\usepackage{csquotes}
\usepackage{amsmath} 

\usepackage{amssymb}
\begin{document}

\title{\model: Dynamic Causality-Aware Root Cause Analysis for Distributed Microservices}

\author{Songhan Zhang}
\affiliation{%
  \institution{The Chinese University of Hong Kong, Shenzhen}
  \city{Shenzhen}
  \country{China}
}
\email{songhanzhang@link.cuhk.edu.cn}

\author{Aoyang Fang}
\affiliation{%
  \institution{The Chinese University of Hong Kong, Shenzhen}
  \city{Shenzhen}
  \country{China}
}
\email{aoyangfang@link.cuhk.edu.cn}

\author{Yifan Yang}
\affiliation{%
  \institution{The Chinese University of Hong Kong, Shenzhen}
  \city{Shenzhen}
  \country{China}
}
\email{yifanyang6@link.cuhk.edu.cn}

\author{Ruiyi Cheng}
\affiliation{%
  \institution{The Chinese University of Hong Kong, Shenzhen}
  \city{Shenzhen}
  \country{China}
}
\email{121020031@link.cuhk.edu.cn}

\author{Xiaoying Tang}
\affiliation{%
  \institution{The Chinese University of Hong Kong, Shenzhen}
  \city{Shenzhen}
  \country{China}
}
\email{tangxiaoying@cuhk.edu.cn}

\author{Pinjia He}
\authornote{Corresponding author.}
\affiliation{%
  \institution{The Chinese University of Hong Kong, Shenzhen}
  \city{Shenzhen}
  \country{China}
}
\email{hepinjia@cuhk.edu.cn}

\renewcommand{\shortauthors}{Songhan Zhang, et al.}

\begin{abstract}
Cloud-native microservices enable rapid iteration and scalable deployment but also create complex, fast-evolving dependencies that challenge reliable diagnosis. Existing root cause analysis (RCA) approaches, even with multi-modal fusion of logs, traces, and metrics, remain limited in capturing dynamic behaviors and shifting service relationships. Three critical challenges persist: \textit{(i) inadequate modeling of cascading fault propagation, (ii) vulnerability to noise interference and concept drift in normal service behavior, and (iii) over-reliance on service deviation intensity that obscures true root causes.}
To address these challenges, we propose \model, a dynamic causality-aware framework for RCA in distributed microservice systems. \model\ unifies multi-modal dynamic signals to capture time-varying spatio-temporal dependencies through interaction-aware representation learning.
It further introduces a dynamic contrastive mechanism to disentangle true fault indicators from contextual noise and adopts a causal-prioritized pairwise ranking objective to explicitly optimize causal attribution.
Comprehensive evaluations on public benchmarks demonstrate that \model\ consistently surpasses state-of-the-art methods, attaining an average \textit{AC@1} of 0.63 with absolute gains of 0.25–0.46, and delivering both accurate and interpretable diagnoses in highly dynamic microservice environments.
\end{abstract}




\maketitle

\section{Introduction}
\label{sec:introduction}
The rapid evolution of cloud-native applications has driven the adoption of microservice architectures, enabling agile development, scalable deployment, and flexible maintenance \cite{balalaie2016microservices,dragoni2017microservices,wang2018cloudranger}.
However, the inherent dynamism of microservices—frequent scaling, shifting invocation relationships, and evolving topologies—poses major reliability challenges~\cite{addeen2019dynamic}.
Dynamic behaviors often exacerbate system fragility: auto-scaling may amplify transient faults into cascading failures; continuous deployment introduces elusive ``ghost'' failures; and adaptive routing across ephemeral instances causes unpredictable latency spikes.
Ensuring reliability in such environments demands advanced diagnostic techniques.
In this regard, Root Cause Analysis (RCA) serves as a cornerstone, guiding system remediation by identifying the fundamental sources of failure~\cite{zhang2024failure}.

\begin{figure}[h]
  \centering
  \includegraphics[width=0.49\textwidth]{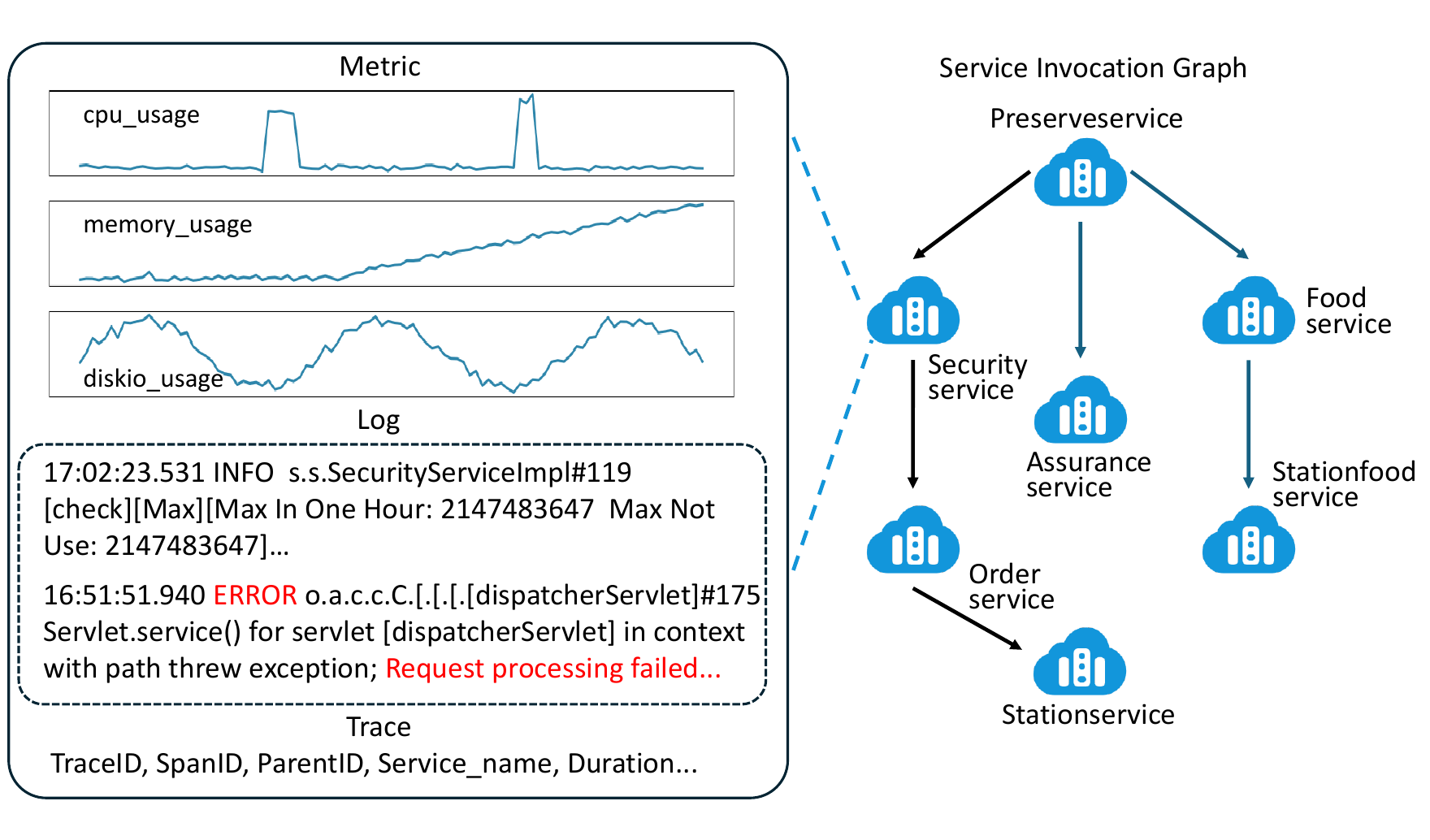}
  \caption{Multi-Modal Monitoring Data and a Service Invocation Graph In a Microservice System. The Left Part Illustrates the Forms of Metrics, Logs, and Traces, Whereas The Right Part Depicts An Example of Service Invocation Graph.}
  \label{fig1:monitoring data}
\end{figure}
As shown in Figure~\ref{fig1:monitoring data}, microservice reliability is typically monitored through three data modalities (metrics, logs, and traces) which form the foundation of RCA.
Meanwhile, the figure also presents a service invocation graph that illustrates the call relationships among different microservices, helping to trace how faults might propagate across these services.
Traditional RCA methods rely on a single modality (e.g., metrics~\cite{pham2024baro,zhang2021cloudrca,ma2020automap,li2022actionable,wang2023interdependent,zhang2024illuminating,somashekar2024gamma}, traces~\cite{ding2023tracediag,gan2023sleuth,wang2023incremental,zhang2024trace,yu2021microrank,lin2018microscope}, or logs~\cite{du2017deeplog,kobayashi2017mining,li2020swisslog}), overlooking the holistic system state.
Each modality has blind spots: resource-level faults (e.g., CPU contention) leave few records in logs, while configuration and logic bugs rarely appear in metrics~\cite{lee2023eadro,wang2024mrca}.
To overcome this, recent studies explore multi-modal fusion RCA that integrates heterogeneous monitoring data to provide unified system insights~\cite{lee2023eadro,yu2023nezha,sun2024art,zhang2023robust,wang2024mrca,zheng2024mulan,zhu2024hemirca,zhang2024no,sun2025interpretable}.
Nevertheless, existing multi-modal RCA approaches still struggle to cope with the dynamic nature of microservices, leaving three key challenges unaddressed.

\textbf{Challenge 1: Inadequate handling of cascading fault propagation.} 
Microservices form dynamic dependency networks where a fault in one service can cascade through inter-service interactions, masking the true root cause behind secondary anomalies \cite{cloudzero2025statistics}. Existing RCA approaches remain limited in addressing this issue due to both modeling and data deficiencies.
(1) \textbf{Modeling limitations.} \textit{Existing RCA methods insufficiently capture system-level interactions: Most rely on shallow or static representations of service interactions, neglecting their temporal evolution and causal dynamics.} For instance, Graph Attention Networks (GAT)-based \cite{velivckovic2017graph} methods such as Eadro \cite{lee2023eadro} emphasize services' fault symptom aggregation rather than modeling causal propagation, and typically treat invocation links as fixed binary relations rather than real-world dependencies evolve as the system state changes.
(2) \textbf{Dataset deficiencies.}
Realistic fault propagation requires both \textit{system-interaction metrics} (e.g., latency, status code in trace) and \textit{service-internal data} (e.g., metrics, logs, traces)~\cite{li2022causal}. 
However, an empirical study claims that \textit{most existing public datasets (Except RCAEval-RE2-OB \cite{pham2025rcaeval}) suffer from incomplete data collection, often missing one or both modalities critical for modeling cascading faults }\cite{fang2025empiricalstudysotarca}.
This incompleteness prevents existing methods from effectively simulating or evaluating real-world fault propagation behaviors.

\textbf{Challenge 2: Challenges of noise interference and concept drift caused by normal fluctuations.}
Dynamic fluctuations in normal microservice operations (e.g., daily peak-hour traffic spikes, stochastic inter-service interactions, monitoring noise) often introduce metric deviations that resemble fault signals. 
Moreover, continuous system evolution (business growth, version updates, architectural adjustments) triggers concept drift, where the statistical distribution of normal behaviors shifts over time and gradually diverges from the models’ learned representations \cite{addeen2019dynamic,bierzynski2019supporting}. 
However, most existing multi-modal RCA methods overlook the essential need to distinguish normal fluctuations from genuine fault-induced deviations under evolving system dynamics. As a result, they tend to conflate normal fluctuations with abnormal behaviors, leading to unstable and unreliable diagnosis in dynamic microservice environments \cite{lee2023eadro,duan2025famos,wang2024mrca}.


\textbf{Challenge 3: Misalignment between service deviation intensity and root-cause relevance.}
We define \textit{service deviation intensity} as the magnitude of deviation between a service’s observed metrics and its normal operating baseline.
Most existing models effectively equate root cause analysis with service deviation intensity ranking, focusing solely on per-service anomaly scores and thereby prioritizing those with the highest deviation intensity \cite{sun2024art,zhang2023robust,yu2023nezha,pham2024baro}.
However, \textit{this intensity is not necessarily indicative of a service’s likelihood of being the true root cause, } since fault propagation and inherent system variability often cause downstream or mission-critical services to exhibit stronger fault symptoms than the actual root cause service.
This overemphasis leads to root cause misattribution, where services with higher deviation intensity are mistakenly prioritized over the real root cause, especially when the true root cause exhibits weaker abnormal deviations than its affected counterparts.

To tackle these challenges, we propose \textbf{\model}, a dynamic-aware framework for root cause analysis in microservice systems.
The design of \model\ is driven by three core, layered insights.
\textit{(1) The essence of RCA lies in understanding fault propagation patterns, not merely identifying isolated anomalies. }
To this end, \model\ introduces a system-interaction-aware representation learner that synergizes a Transformer with a novel Graph Attention Network (GAT) to cohesively model the spatio-temporal "fingerprint" of a fault, enabling precise traceability.
\textit{(2) The definition of normality in microservice systems evolves with context.}
To disentangle the true fault signal from normal system fluctuations, \model\ pioneers a dynamic contrastive representation mechanism.
By contrasting a service's current state against its own historical normal baseline, it effectively identifies the semantic shift driven by the root cause, thereby filtering out environmental noise.
\textit{(3) The service exhibiting the largest deviation is not necessarily the true root cause.}
To overcome misranking caused by an over-reliance on service deviation intensity, \model\ incorporates a root-cause-prioritized pairwise ranking objective that embeds causal priors during training, compelling the model to rank the actual root cause above its affected services.
This design enhances RCA accuracy, particularly when the root cause shows only subtle deviations.

To ensure a fair and comprehensive evaluation, we compare \model\ with state-of-the-art baselines on two datasets: RCAEval-RE2-OB \cite{pham2025rcaeval}, a well-established benchmark, and a newly released large-scale dataset \cite{fang2025empiricalstudysotarca}—currently the largest RCA benchmark by fault case count, covering more diverse fault types and propagation scenarios.
Extensive experiments show that \model\ consistently surpasses all baselines, achieving an average \textit{AC@1} of 0.63 and absolute improvements of 0.25–0.46, while delivering accurate and interpretable diagnoses in dynamic microservice systems.

The main contributions of this work are summarized as follows.
\begin{itemize}
    \item We identify three critical challenges for dynamic RCA in microservices, cascading fault propagation, normal fluctuation confusion, and deviation–cause misalignment, analyzing why existing methods fall short.
    \item We propose \model, which integrates a dynamic feature alignment module and system-interaction-aware representation learner to capture temporal dynamics and system-wide causal interactions.  
    \item We introduce a dynamic contrastive mechanism and a root-cause-prioritized ranking module, unified under a \textit{Causal Representation Discrimination (CRD)} framework in our methodological design, to separate true anomalies from normal deviations and mitigate deviation-intensity bias.
    \item Experiments on public datasets show that \model\ achieves an average \textit{AC@1} of 0.63, outperforming baseline methods by 0.25–0.46.
\end{itemize}

\section{Motivation}
\label{sec:motivation}
\begin{figure*}[t]
    \centering
    \begin{subfigure}[t]{0.6\textwidth}
        \centering
        \includegraphics[width=\textwidth]{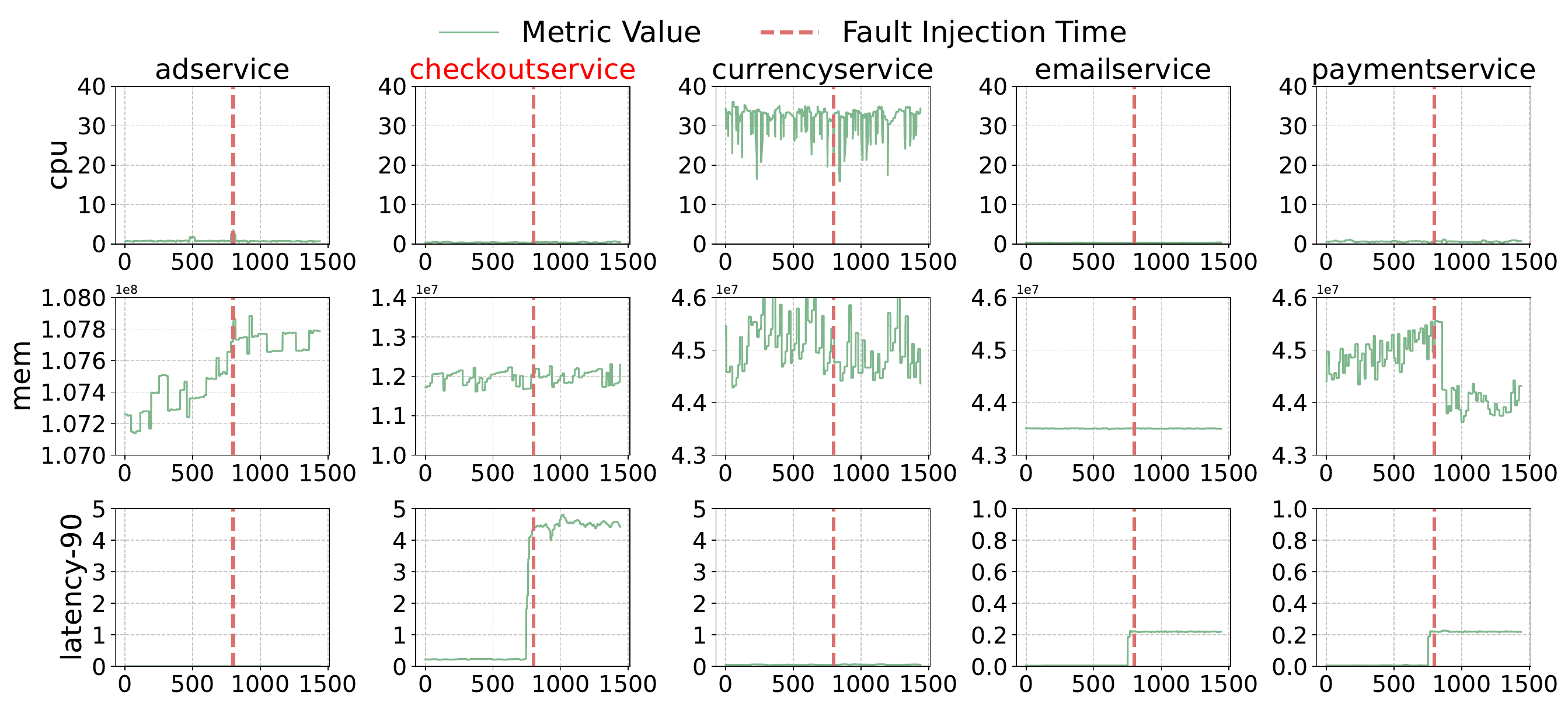}
        \caption{Service-Level Metric value Before and After Fault Injection.}
        \label{Fig2:subfig1}
    \end{subfigure}
    \hspace{3pt}
    \begin{subfigure}[t]{0.37\textwidth}
        \centering
        \includegraphics[width=\textwidth]{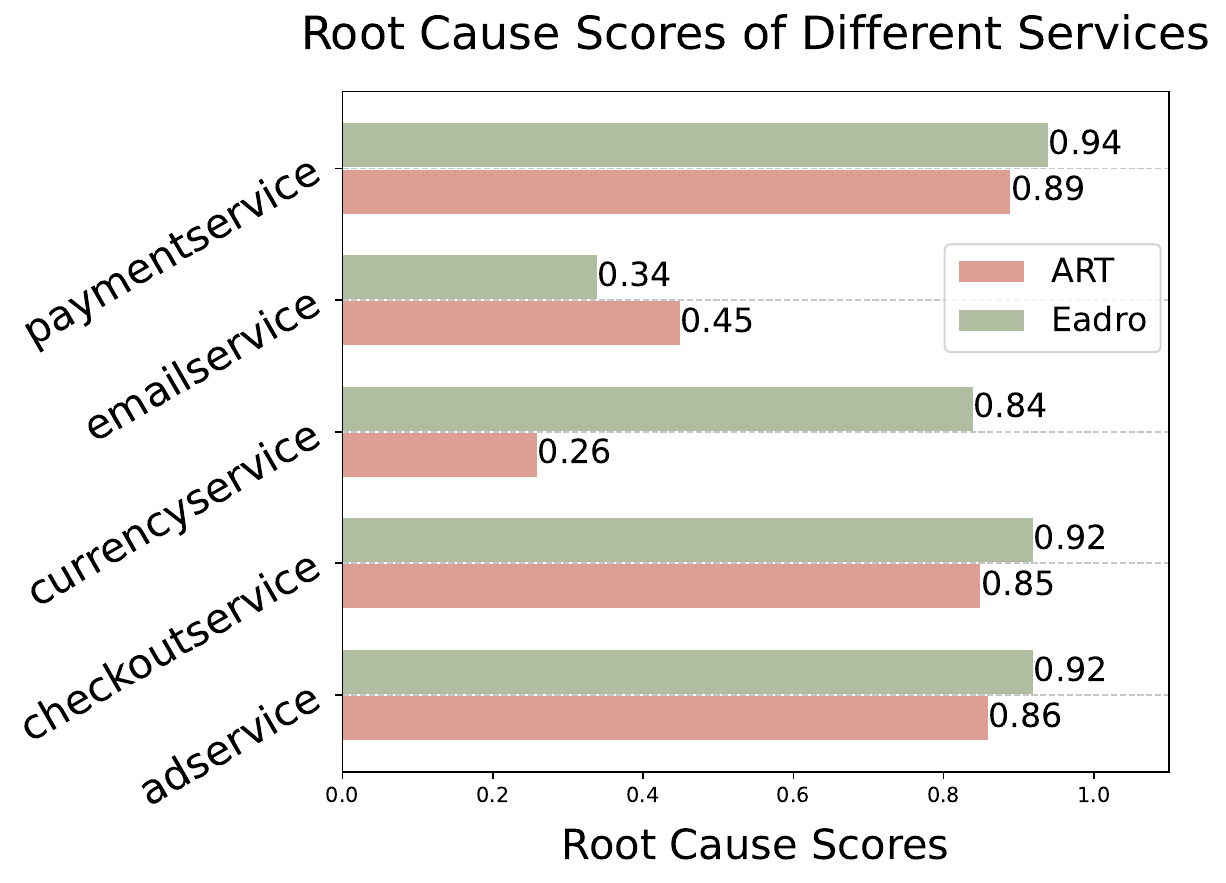}
        \caption{Existing Approaches’ Root Cause Scoring for Different Services in the Target Case.}
        \label{Fig2:subfig2}
    \end{subfigure}
    \caption{Network Delay Fault Case in RCAEVAL-RE2-OB: Metrics and Root Cause Scores. (a) CPU of \texttt{CurrencyService} Has Same Fluctuation Pattern in Abnormal/Normal Periods; \texttt{Paymentservice} and \texttt{Adservice} Memory Shows More Deviation After \texttt{Checkoutservice} Fault Injection. (b) Existing Methods’ Root Cause Scores for \texttt{Paymentservice} and \texttt{Adservice} Exceed That for \texttt{Checkoutservice} (Ground Truth).}
    \label{Fig2: motivation_example}
\end{figure*}
To bridge the identified research gaps, we revisit two pivotal questions that remain unresolved by current state-of-the-art (SOTA) RCA methods. These questions directly correspond to the three challenges discussed in Section~\ref{sec:introduction}: Q1 relates to incomplete modeling of fault propagation and normal fluctuations in microservices' normal operations (Challenge 1 \& 2), while Q2 addresses the misranking of root causes due to over-reliance on service deviation intensity (Challenge 3).

\subsection{Can Incomplete System Interaction Modeling and normal Fluctuations Undermine Microservice Reliability?}

\textbf{Incomplete system interaction modeling leads to the loss of fault propagation information.} 
In microservice systems, faults typically originate from intra-service anomalies (e.g., CPU saturation, memory leaks) and then propagate through inter-service interactions (e.g., interface latency spikes, 5xx error cascades).
For instance, a CPU saturation in Service A drastically increases its processing time, causing its thread pool to become exhausted. This resource bottleneck then forces its downstream caller, Service B, to wait or timeout, which is observed as a rapid elevation in B’s P90 latency and 5xx status codes.
Therefore, effective RCA methods must capture both intra-service states and inter-service dependencies.
However, most existing approaches pay limited attention to the latter, focusing primarily on aggregating node-level information via graph-based learning on intra-service metrics.
As a result, they often fail to model the causal dependencies and dynamic propagation patterns that are crucial for accurate fault localization \cite{sun2024art,lee2023eadro,zhang2023robust}.

\textbf{Lack of normal–anomaly contrast causes confusion with normal fluctuations.}
Existing multimodal approaches \cite{zhang2023robust,lee2023eadro,wang2024mrca} primarily emphasize end-to-end fault pattern learning, yet they rarely establish an explicit contrast between fault-period characteristics and a service’s normal-state baseline.
As illustrated in Figure~\ref{Fig2:subfig1}, which depicts a network-delay fault from an existing benchmark dataset \cite{pham2025rcaeval}, the fault is injected into the \texttt{checkoutservice}.
However, for the CPU metric, a relatively higher value compared to other services can be observed in the \texttt{currency\-service}, and it also shows the same fluctuations and value level during its normal operation period.
Nevertheless, the \texttt{currencyservice} exhibits a relatively higher and more fluctuating CPU pattern than other services during the fault period, a pattern that consistently persists even under normal conditions.
When a model learns fault-specific patterns without incorporating such normal–anomaly contrast, it risks mistaking these normal fluctuations for fault signatures. 
This misalignment inevitably leads to a higher false positive rate in root cause analysis, as normal system fluctuations are incorrectly attributed to anomalous behavior.

\subsection{Does Overemphasis on Service Deviation Intensity Lead to Root Cause Misranking?}

\textbf{Over-reliance on deviation intensity leads to root cause misranking.}
Most existing RCA methods implicitly assume that a service with a higher deviation intensity is more likely to be the root cause \cite{zhang2023robust,yu2023nezha,chen2014causeinfer}. However, in dynamic microservice environments, fault propagation and inherent fluctuations often amplify anomalies in affected services, making non-root-cause services appear more abnormal than the true root cause. 
We examine this issue through a case study on the RCAEVAL-RE2 benchmark dataset \cite{pham2025rcaeval}, as shown in Figure~\ref{Fig2: motivation_example}. In this network delay fault, the injected root cause \texttt{checkoutservice} experiences only latency degradation (increasing from its normal baseline of ~0.2s to ~4.8s) without any other metric anomaly. In contrast, the downstream \texttt{paymentservice} and \texttt{adservice} exhibit pronounced memory level shifts—approximately an order of magnitude larger than normal—which visually dominate the anomaly patterns despite being secondary effects.

When we applied two state-of-the-art multi-modal RCA methods (ART \cite{sun2024art} and Eadro \cite{lee2023eadro}), both incorrectly ranked these non-root services higher than \texttt{checkoutservice}, as shown in Figure~\ref{Fig2:subfig2}. 
\textbf{This case illustrates that existing models tend to overfit to the magnitude of service deviations while neglecting causal dependencies among services, leading to misranking of the true root cause.}

\section{Methodology}
\label{sec:methodology}

\begin{figure*}[h]
  \centering
  \includegraphics[width=1.\textwidth]{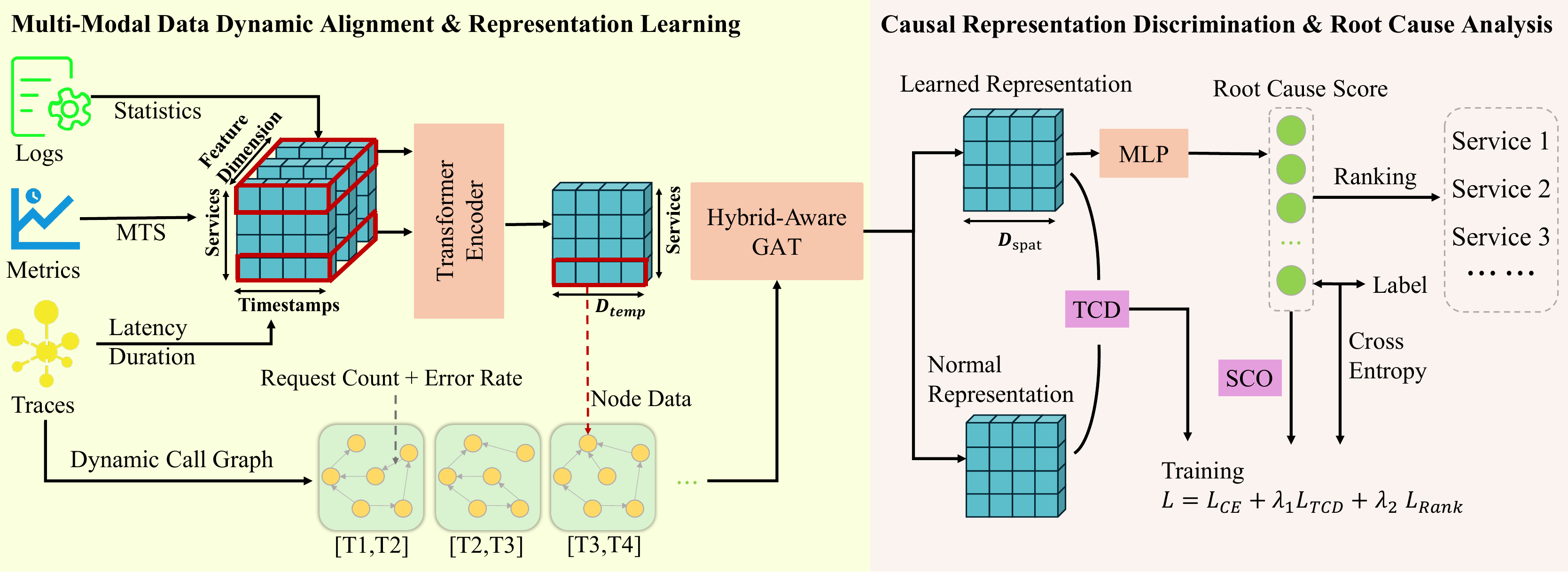}
  \caption{Framework of \model: Multi-Modal Data Dynamic Alignment, System Interaction-Driven Representation Learning, Causal Representation Discrimination with TCD (Temporal Causal Disentanglement) and SCO (Spatial Causal Ordering), as well as Root Cause Analysis.}
  \label{fig3:method}
\end{figure*}

To effectively address the challenges of dynamic microservice dependencies, fault propogation, noise interference, and service deviation intensity over-reliance discussed in Section~\ref{sec:motivation}, we propose \model, a dynamic-aware framework for Root Cause Analysis. As shown in Figure~\ref{fig3:method}, the \model\ framework consists of three primary stages: (i) a \textbf{multi-modal dynamic alignment module} ; (ii) a \textbf{system-interaction-aware representation learning} for fault propagation; and (iii) a \textbf{causal representation discrimination ($\text{CRD}$)} module to enhance robustness against dynamic noise and root cause misranking.

\subsection{Problem Statement}
Following existing RCA methods \cite{lee2023eadro,yu2023nezha,zhang2023robust,sun2024art, jiang2023look,yu2023cmdiagnostor,shan2019diagnosis}, we define the service-level RCA task in dynamic microservice systems as follows: Given an incident triggered at time window $[t, t+T]$, and multi-modal monitoring data of $\mathcal{V}$ microservices over a time window, including metrics,traces, and logs, the goal is to compute a root cause score vector $S^{(t)} \in [0,1]^\mathcal{V}$. Here, $S_i^{(t)}$ quantifies the likelihood that service $i$ is the root cause of the triggered system anomalies at time window $[t, t+T]$. We then rank all services in descending order of their $S_i^{(t)}$ values, with the top-ranked service identified as the primary candidate for the fault’s root cause, enabling precise localization of the source driving the observed anomalies.

\subsection{Multi-modal Data Alignment}
\label{sec:data_model}

In microservice systems, faults propagate through inter-service dependencies. Accurate modeling requires the integration of real-time service behavior (service-internal features) and the time-varying interaction topology (graph structure).

\textbf{Multi-modal Time Series Feature Alignment.} 
Following existing methods \cite{lee2023eadro,sun2024art}, we collect three fundamental modalities of monitoring data (e.g., metrics, logs, traces), transforming them into multivariate time seires (MTS).
Since metrics are inherently time series, we only perform sampling to ensure uniform time intervals across data points. For logs, we extract statistical features such as the counts of events and severity levels at each sampling point to construct time series. For traces, we convert latency into a sequential form. For each service $i \in \mathcal{V}$ at time $t$, we obtain a unified multi-modal feature vector $\mathbf{X}_i^{(t)} \in \mathbb{R}^{D_{\text{metrics}}+D_{\text{logs}}+D_{\text{traces}}}$, where  $D_{\text{feature}}$
is the dimension of the feature.
We then apply min–max normalization \cite{henderi2021comparison} to eliminate scale disparities among different modalities, which facilitates stable gradient updates and accelerates convergence during model training.
Finally,  we segment the MTS into time windows of length $T$, which not only preserves local temporal dependencies but also enables the model to capture evolving patterns within different time spans.
The resulting input is a sequence of feature vectors for each service, $\mathbf{X}^{t} = \{\mathbf{X}_1^{(t:t+T)}, \dots, \mathbf{X}_{|\mathcal{V}|}^{(t:t+T)}\}$.

\textbf{Dynamic Call Graph Construction.}
Existing graph-based RCA methods often rely on static, binary call relationships which fail to capture the real-time systems states. To overcome this, we construct a dynamic call graph $\mathcal{G}_t = (\mathcal{V}, \mathcal{E}_t)$ for every time window $t$.
The adjacency matrix $\mathbf{A}_t$ is defined where the edge weight $e_{i,j}^{(t)}$ from calling service $i$ to called service $j$ is dynamically calculated. 
To better capture system interaction information and fault propagation characteristics, we use a combination of real-time request count and status code-based API error rate to quantify the edge weight, since fault propagation between services often causes abnormal fluctuations in request count (e.g., surges from retries or drops from service halts) and increased error rates (e.g., 5xx or timeouts). The edge weight is defined as: 
\begin{equation}
e_{i,j}^{(t)} = \sigma\left(\alpha \cdot \text{Norm}(C_{i,j}^{(t)}) + (1-\alpha) \cdot \text{Norm}(R_{i,j}^{(t)})\right)
\end{equation}
where $C_{i,j}^{(t)}$ and $R_{i,j}^{(t)}$ are the normalized request count and error rate for the interaction from $i$ to $j$ at time $t$, respectively, $\sigma(\cdot)$ is the sigmoid function, and $\alpha \in [0, 1]$ is a hyperparameter balancing the two modalities. This dynamic edge weighting allows \model\ to quantify the real-time interaction intensity and risk, offering a superior representation for subsequent fault propagation modeling compared to binary or static graphs.

\subsection{System-Interaction-Aware Representation Learning}
\label{sec:representation_learning}
To better capture the real-time state of the entire system, this stage learns the spatio-temporal embedding $\mathbf{H}_i^{s,t}$ through a two-step process: first encoding service-internal dynamic features (i.e., preprocessed MTS data for each service) using a Transformer encoder \cite{vaswani2017attention} to model temporal dynamics, then integrating these service-level embeddings into graph nodes and capturing inter-service interaction patterns (including call dependencies, request flows, and fault propagation information) via our proposed Hybrid-Aware GAT to model spatial relationships.

\subsubsection{Temporal Feature Encoding}
Transformers are widely used for sequence and temporal modeling, as they efficiently capture long-range dependencies and complex patterns in time-series data while enabling parallel computation across time steps, making them ideal for modeling the high-dimensional multivariate temporal features \cite{vaswani2017attention,zhou2021informer,tuli2022tranad}.
To capture services' dynamic internal states, We apply a Transformer Encoder to process the look-back sequence of feature vectors $\mathbf{X}_i = \{\mathbf{X}_i^{(t)}, \dots, \mathbf{X}_i^{(t+T-1)}\} (i \in \mathcal{V})$. 
The encoder first projects the input features into query ($Q$), key ($K$), and value ($V$) matrices via linear transformations: 

\begin{equation}
\mathbf{Q}_i = \mathbf{X}_i \mathbf{W}_Q, \quad \mathbf{K}_i = \mathbf{X}_i \mathbf{W}_K, \quad \mathbf{V}_i = \mathbf{X}_i \mathbf{W}_V
\end{equation}

where $\mathbf{W}_Q, \mathbf{W}_K,\mathbf{W}_V \in \mathbb{R}^{F \times D_{\text{k}}}$ are learnable parameter matrices, and $D_{\text{k}}$ is the dimension of the query/key space. Then, multi-head self-attention is then computed across the time steps to model temporal relationships:

\begin{equation}
\text{Attention}(\mathbf{Q}_i, \mathbf{K}_i, \mathbf{V}_i) = \text{softmax}\left(\frac{\mathbf{Q}_i \mathbf{K}_i^\top}{\sqrt{D_k}}\right) \mathbf{V}_i
\end{equation}

After concatenating outputs from multiple attention heads and applying a feed-forward network with residual connections and layer normalization, the Transformer Encoder generates a time-aware embedding $\mathbf{H}_{i}^{(t)} \in \mathbb{R}^{D_{\text{temp}}}$ for each service $i$ at time $t$:
\begin{equation}
\mathbf{H}_i^{(t)} = \text{TransformerEncoder}(\mathbf{X}_i^{(t)}, \dots, \mathbf{X}_i^{(t+T-1)})
\end{equation}

\subsubsection{Hybrid-Aware Graph Attention Network (H-GAT)}
Standard Graph Attention Networks (GATs) \cite{velivckovic2017graph} model inter-service relationships through attention coefficients learned automatically from node features, enabling message passing and information aggregation across connected services. While effective for capturing general relational patterns, this approach aggregates service-internal states based solely on learned similarity, without explicitly accounting for the causal propagation patterns and characteristic behaviors observed in fault cascades.
Specifically, standard GATs fail to leverage the rich diagnostic signals in monitoring data (e.g., inter-service interface metrics) that indicate how faults actually propagate through a microservice system.

To address this limitation, \model\ introduces a novel HA-GAT with a dual-mechanism layer that combines two complementary aggregation strategies: one focusing on service-internal state relationships, and another explicitly modeling fault propagation causality.
The first mechanism employs standard GAT attention to capture intrinsic feature-based relationships between services. For the$k$-th layer, attention coefficients \(\alpha_{i,j}^{(k)}\)are computed based on the similarity of transformed node features:

\begin{equation}
\alpha_{i,j}^{(k)} = \frac{\exp\left(\text{LeakyReLU}\left(\mathbf{a}^T \cdot \left[\mathbf{W}\mathbf{H}{i}^{(k-1)} || \mathbf{W}\mathbf{H}{j}^{(k-1)}\right]\right)\right)}{\sum_{k \in \mathcal{N}(i)} \exp\left(\text{LeakyReLU}\left(\mathbf{a}^T \cdot \left[\mathbf{W}\mathbf{H}{i}^{(k-1)} || \mathbf{W}\mathbf{H}{k}^{(k-1)}\right]\right)\right)}
\end{equation}
where \(\mathbf{W}\) is a learnable transformation matrix, \(\mathbf{a}\) is an attention vector, and \(\mathcal{N}(i)\) denotes the neighbors of service $i$. This produces feature-aware message passing weights that capture how service-internal states relate to one another.

The second mechanism integrates the pre-calculated dynamic edge weights $e_{i,j}^{(t)}$ (derived from real-time request counts and error rates) to model causal propagation intensity. These weights regulate the message passing process, ensuring that information flow reflects actual fault propagation patterns:

\begin{equation}
\mathbf{H}{i}^{(k)} = \sigma \left(\sum{j \in \mathcal{N}(i)} \alpha_{i,j}^{(k)} \cdot e_{i,j}^{(t)} \cdot \mathbf{W}^{(k)} \mathbf{H}_{j}^{(k-1)} \right)
\end{equation}

For the final output layer ($K$-th layer), the process yields a set of service-level embeddings \(\mathbf{H}^{(s,t)} = \{\mathbf{H}_1^{(K)}, \mathbf{H}_2^{(K)}, \dots, \mathbf{H}_N^{(K)}\} \in \mathbb{R}^{N \times D_{\text{spat}}} \) (with\(D_{\text{spat}}\) as the representation feature dimension), where each row \(\mathbf{H}_i^{(K)}\) represents the integrated spatio-temporal embedding of service $i$.
Finally, we map it to a \((\mathcal{V}, 1)\) vector as the root cause score via a multi-layer perceptron (MLP) \cite{lecun2015deep}, where each entry quantifies the probability of the corresponding service being the root cause, enabling a ranking of root cause likelihoods. 

\begin{equation}
\mathbf{S}^{(t)} = \sigma\left(\mathbf{W}_2 \cdot \text{ReLU}\left(\mathbf{W}_1 \cdot \mathbf{H}^{(s,t)} + \mathbf{b}_1\right) + \mathbf{b}_2\right)
\end{equation}
where $\mathbf{W}_1, \mathbf{W}_2$ are learnable weight matrices, $\mathbf{b}_1, \mathbf{b}_2$  are bias terms, $\sigma $ denotes the Sigmoid activation function (constraining scores to the range $[0,1]$). 
By fusing feature-based attention with causal propagation weights, HA-GAT achieves two key advantages: (1) it more effectively captures the system interaction patterns by grounding message passing in real-world fault propagation signals, and (2) it enhances model interpretability by aligning attention with observable system metrics (request flows and error rates) that engineers use to diagnose issues.
$\mathbf{S}^{(t)}$ is the final root cause score vector with each entry \(S_i^{(t)}\) representing the likelihood that service $i$ is the root cause of the observed system anomalies at time $t$.

\subsection{Causal Representation Discrimination ($\text{CRD}$)}
\label{sec:crd}
In order to address the vulnerability of RCA models to normal fluctuations, and root cause misranking caused by excessive focus on service internal states, \model\ introduces a \textbf{Causal Representation Discrimination (CRD)} module. CRD integrates the dynamic contrastive mechanism (temporal dimension) and the root-cause-prioritized ranking mechanism (spatial dimension), enforcing complementary constraints on the learned representations $\mathbf{H}$ and output scores $S$. This ensures that the identified root cause possesses genuine causal precedence, robust against dynamic fluctuations and misranking.

\subsubsection{Temporal Causal Disentanglement ($\text{TCD}$)}
\label{sec:tcd}

To address noise interference and concept drift, the $\text{TCD}$ focuses on disentangling the root-cause-induced shifts from normal fluctuations by using a contrastive mechanism on the latent feature space $\mathbf{H}$.

We compare the anomalous state representation $\mathbf{H}_i^{\text{anom}}$ (current $\mathbf{H}_{i}^{s, t}$) against its normal state representation $\mathbf{H}_i^{\text{norm}}$ to quantify their dissimilarity.
The key principle is to maximize this dissimilarity for the true root cause $r$, while simultaneously minimizing it for non-root causes $i$. This is formalized as the TCD loss:
\begin{equation}
L_{TCD} = \sum_{i \neq r} \max(0, \delta - \cos(\mathbf{H}_r^{\text{anom}}, \mathbf{H}_r^{\text{norm}})+\cos(\mathbf{H}_i^{\text{anom}}, \mathbf{H}_i^{\text{norm}}))
\end{equation}
where $\cos(\cdot, \cdot)$ is the cosine similarity, and $\delta$ are margin boundaries. 

By embedding these contrastive constraints into the optimization, TCD magnifies the temporal divergence between pre- and post-fault embeddings of true root-cause services, while suppressing such divergence for non-root ones. This temporal disentanglement effectively filters out normal fluctuations, thereby reducing false positives and enabling more accurate attribution of anomalies to genuine root causes.

\subsubsection{Spatial Causal Ordering ($\text{SCO}$)}
\label{sec:sco}

To mitigate root cause misidentification caused by over-reliance on service-internal states, the Spatial Causal Ordering ($\text{SCO}$) module leverages the spatial dependency structure of microservices to impose causal ranking constraints on the prediction scores $S_i$.

The $\text{SCO}$ enforces that the true root cause service's score $S_r$ must consistently rank higher than the scores of all its affected downstream and upstream services $i \in \mathcal{P}(r)$ ( $\mathcal{P}(r)$ is a set of affected downstream and upstream services of $r$). 
This ordering is translated into a pairwise ranking objective: 

\begin{equation}
L_{\text{rank}} =  \sum_{i \in \mathcal{P}(r)} \max(0, m - (S_r - S_i))
\end{equation}
where $m > 0$ is a margin hyperparameter. 

By integrating this ranking loss, SCO explicitly encodes the causal ordering implied by fault propagation across the service topology. As a result, the framework systematically distinguishes true sources from their propagated effects, thereby improving the robustness and interpretability of root-cause ranking.. 

\subsection{Model Training and Inference}
\label{sec:optimization}

The overall objective function of \model\ is a weighted combination of the standard Cross-Entropy loss ($L_{CE}$) for classification and the two auxiliary loss terms from the $\text{CRD}$ module:

\begin{equation}
L_{\text{total}} = L_{CE} + \lambda_{1} L_{TCD} + \lambda_{\text{2}} L_{\text{rank}}
\end{equation}

where $\lambda_{1}$ and $\lambda_{\text{2}}$ are weighting coefficients.

During \textbf{Inference}, a new time window's multi-modal data is processed by \model\ to obtain the final anomaly intensity scores $\mathbf{S}$. The services are ranked in descending order based on $S_i$, and the service with the highest score (Rank 1) is identified as the predicted root cause.

\section{Experiment}
\begin{table}[t!]
\centering
\caption{Statistics of Datasets. Svc: Services, F. C. Svc: Fault-Covered Services, Train: Train Set, Test: Test Set.}
\label{Tab0: dataset}
\resizebox{0.47\textwidth}{!}{%
\begin{tabular}{l|rrrrrr}
\toprule
Dataset & \# Svc & \# Fault & \# Type & \# F. C. Svc & \# Train & \# Test \\ \midrule
D1      & 12     & 90       & 6       & 5            & 60           & 30          \\
D2      & 50     & 1430      & 25      & 42           & 1145          & 285          \\ \bottomrule
\end{tabular}
}
\end{table}
\begin{table*}[t!]
\centering
\caption{Overall Performance Comparison of \model\ with SOTA Baseline Methods. Evaluations are conducted on two datasets (D1, D2) using five metrics: Top-k Accuracy (AC@1, AC@3, AC@5), Average Precision@5 (Avg@5), and Mean Reciprocal Rank (MRR). The best performance in each metric is highlighted in bold.}
\label{Tab1: Overall performance}
\scriptsize
\resizebox{0.95\textwidth}{!}{%
\begin{tabular}{l|rrrrr|rrrrr}
\toprule
\multicolumn{1}{c|}{\multirow{2}{*}{\centering \textbf{Method}}} & \multicolumn{5}{c|}{\textbf{D1}} & \multicolumn{5}{c}{\textbf{D2}} \\  \cmidrule(lr){2-6} \cmidrule(lr){7-11}
\multicolumn{1}{c|}{} & \multicolumn{1}{c}{\textbf{AC@1}} & \multicolumn{1}{c}{\textbf{AC@3}} & \multicolumn{1}{c}{\textbf{AC@5}} & \multicolumn{1}{c}{\textbf{Avg@5}} & \multicolumn{1}{c|}{\textbf{MRR}} & \multicolumn{1}{c}{\textbf{AC@1}} & \multicolumn{1}{c}{\textbf{AC@3}} & \multicolumn{1}{c}{\textbf{AC@5}} & \multicolumn{1}{c}{\textbf{Avg@5}} & \multicolumn{1}{c}{\textbf{MRR}} \\ 
\midrule
MicroRank & 0.444   & 0.644   & 0.722    & 0.607   & 0.570  & 0.013   & 0.092   & 0.118   & 0.066   & 0.051    \\
Baro      & 0.144   & 0.933   & 1.000    & 0.776   & 0.537  & 0.263   & 0.526   & 0.605   & 0.476   & 0.410    \\
Nezha    & 0.078   & 0.233   & 0.344    & 0.227   & 0.173  & 0.092   & 0.171   & 0.243   & 0.169   & 0.264    \\
Eadro     & 0.459   & 0.546   & 0.547    & 0.526   & 0.501  & 0.301   & 0.511   & 0.671   & 0.504   & 0.467    \\
ART    & 0.196    & 0.578  & 1.000    & 0.593    & 0.452   & 0.176   & 0.232   & 0.250   & 0.233   & 0.231    \\
\textbf{\model} & \textbf{0.769} & \textbf{0.980} & \textbf{1.000} & \textbf{0.937} & \textbf{0.873} & \textbf{0.481} & \textbf{0.696} & \textbf{0.819} & \textbf{0.680} & \textbf{0.626}    \\ \bottomrule
\end{tabular}
}
\end{table*}

In this section, we evaluate the performance of \model\ by focusing on addressing the following research questions:

\textbf{RQ1:} How does \model\ perform in root cause analysis when compared with state-of-the-art baselines?

\textbf{RQ2:} What is the contribution of each component within \model\ to its overall performance?

\textbf{RQ3:} How do hyperparameters affect the performance of \model?

\subsection{Experiment Setup}

\textbf{Datasets.}
To evaluate the performance of \model, we conduct extensive experiments on two real-world microservice system datasets, denoted as D1 and D2.
\begin{itemize}
    \item D1 is a multi-modal fault dataset from the Online Boutique system \cite{OnlineBoutique} (collected by RCAEVAL-RE2 dataset \cite{pham2025rcaeval}). It includes 90 fault cases, 6 resource-related fault types across 5 services (e.g., CPU, Memory, Network Loss). 
    \item D2 is a multi-modal fault dataset from \cite{fang2025empiricalstudysotarca}, currently the largest-scale public RCA dataset for microservices to our knowledge. Built on the 50-service Train-Ticket system \cite{trainticket} (deployed on Kubernetes \cite{kubernetes}), it contains 1,430 fault cases—covering diverse fault types and explicit fault propagation characteristics. This raises RCA difficulty, enabling comprehensive evaluation of RCA methods.
\end{itemize}

For both datasets, when splitting into training and test sets, we ensured the uniform distribution of case types and injected services. Each fault case was uniquely assigned to either the training or test set to avoid data leakage. Detailed statistics are presented in Table~\ref{Tab0: dataset}.

\textbf{Baseline Methods.}
To comprehensively validate \model’s superiority, we select five state-of-the-art (SOTA) RCA methods as baselines, including two unimodal methods and three multi-modal methods:
\begin{itemize}
    \item \textbf{MicroRank \cite{yu2021microrank}:} A trace-based method that analyze normal and abnormal traces to rank potential root cause.
    \item \textbf{Baro \cite{pham2024baro}:} A metric-based method that leverages the Multivariate Bayesian Online Change Point Detection technique to pinpoint root cause.
    \item \textbf{Eadro \cite{lee2023eadro}:} An end-to-end framework that integrates anomaly detection and root cause analysis based on multi-modal data.
    \item \textbf{Nezha \cite{yu2023nezha}:} An multi-modal data-driven method that pinpoints root causes at the code region and resource type level.
    \item \textbf{ART \cite{sun2024art}:} An unsupervised method that combines anomaly detection, failure triage, and root cause analysis.
\end{itemize}

\textbf{Evaluation Metrics.}
For quantitative evaluation, we adopt five widely-used metrics: \textit{AC@1}, \textit{AC@3}, \textit{AC@5}, \textit{Avg@5}, and \textit{MRR}.
\textit{AC@k} (Accuracy@k) Proportion of cases where the true root cause is in top-k candidates: $\text{AC}@k = \frac{1}{N} \sum_{i=1}^N \mathbb{I}(r_i \in \text{Top-}k(S_i))$, where $N$ is the total number of failure cases, $r_{i}$ is the true root cause of the $i$-th case, $S_{i}$ is the ranked list of candidates, and $\mathbb{I}(\cdot)$ is the indicator function. \textit{Avg@5} is the average rank of true root causes within top-5: $\text{Avg}@5 = \frac{1}{5} \sum_{k=1}^5 \text{AC@}k$. 
\textit{MRR} (Mean Reciprocal Rank) is the average reciprocal of true root cause ranks: $\text{MRR} = \frac{1}{N} \sum_{i=1}^N \frac{1}{\text{rank}(r_i, S_i)}$. Better RCA performance is indicated by higher \textit{AC@k}, \textit{Avg@5} and \textit{MRR}.

\subsection{RQ1: Overall Performance}

\textbf{Overall, DyRCA significantly outperforms all baseline methods across both datasets.} As show in Table~\ref{Tab1: Overall performance}, DyRCA achieves an \textit{AC@1} of 0.769, an \textit{Avg@5} of 0.937, and an \textit{MRR} of 0.873 on D1, which are 0.505, 0.391, and 0.425 higher than the baseline averages, respectively. On D2, the improvements remain substantial, with DyRCA surpassing the baseline averages by 0.312 (\textit{AC@1}), 0.391 (\textit{Avg@5}), and 0.341 (\textit{MRR}). Even when compared against the strongest baseline method, Eadro, DyRCA demonstrates clear superiority: DyRCA’s \textit{AC@1} is higher by 0.310 on D1 and by 0.180 on D2. These results highlight DyRCA’s robust capability in precisely identifying the root cause service. We also observe that the performance on D1 is generally higher than on D2, with three baseline methods reaching perfect accuracy at \textit{AC@5} on D1. This discrepancy can be attributed to the increased diversity and complexity of fault types and patterns in D2, as well as the larger number of services in TrainTicket compared to OnlineBique, which collectively make the RCA task more challenging.

\textbf{We further analyze why baseline methods fall short compared to DyRCA.} MicroRank and Baro are single-modal methods (trace-based and metric-based, respectively), which restricts their ability to capture the system’s holistic state and inevitably leads to blind spots in diagnosis. ART, being an unsupervised method, does not leverage label information and instead assumes that the service with the strongest fault symptoms is the root cause. However, this assumption often fails in practice, as the most prominent fault symptom may stem from normal fluctuations or fault propagation rather than the root cause itself. Nezha, although multi-modal, relies on statistical modeling and thus struggles to adaptively learn complex failure patterns compared to learning-based approaches. Eadro, as a supervised multi-modal method, demonstrates competitive performance but is limited by its reliance on static graph structures, neglecting the dynamic interactions and evolving dependencies in microservice systems. Consequently, its effectiveness remains suboptimal compared to \model, which explicitly incorporates dynamic causal modeling.

\subsection{RQ2: Ablation Study}
\begin{table}[t!]
\centering
\caption{Ablation Study of \model: Performance on D1 and D2 Using Metrics AC@1, AC@3, and Avg@5. The study verifies the effectiveness of key components (TCD, SCO, H-GAT) and the impact of dynamic modeling.}
\label{Tab2: Ablation Study}
\scriptsize
\setlength{\tabcolsep}{3pt}
\resizebox{0.48\textwidth}{!}{%
\begin{tabular}{l|rrr|rrr} 
\toprule
\multicolumn{1}{c|}{\multirow{2}{*}{\textbf{Method}}} & \multicolumn{3}{c|}{\textbf{D1}} & \multicolumn{3}{c}{\textbf{D2}} \\  
\cmidrule(lr){2-4}  
\cmidrule(lr){5-7}  
& \textbf{AC@1} & \textbf{AC@3} & \textbf{Avg@5} & \textbf{AC@1} & \textbf{AC@3} & \textbf{Avg@5} \\  
\midrule
\textbf{\model} & \textbf{0.769} & \textbf{0.980} & \textbf{0.937} & \textbf{0.481} & \textbf{0.696} & \textbf{0.680} \\
w/o TCD   & 0.743 & 0.962 & 0.918 & 0.454 & 0.678 & 0.662 \\
w/o SCO   & 0.737 & 0.953 & 0.912 & 0.447 & 0.669 & 0.667 \\
w/o H-GAT & 0.697 & 0.955 & 0.911 & 0.401 & 0.681 & 0.645 \\
w Static  & 0.669 & 0.956 & 0.905 & 0.395 & 0.658 & 0.632 \\ \bottomrule
\end{tabular}
}
\end{table}
To answer RQ2, we conduct an ablation study to investigate the contributions of different components in our proposed model by selectively removing them and evaluating performance. Specifically, we remove the temporal causal disentanglement (TCD) module (denoted as w/o TCD) to test whether modeling temporal causal information improves robustness to normal fluctuations. We remove the spatial causal ordering (SCO) module (w/o SCO) to verify whether our pairwise ranking loss alleviates root-cause misidentification. We further replace the proposed hybrid-aware GAT with a vanilla GAT (w/o H-GAT) to test the benefits of our novel interaction modeling mechanism. Finally, we replace dynamic graph construction with static graphs (w Static) to evaluate the effect of capturing microservice dynamics.

\textbf{Overall, all components contribute effectively to the performance of our model.} As shown in Table~\ref{Tab2: Ablation Study}, all ablated variants underperform compared to the full model. Specifically, the \textit{AC@1} of ablated variants decreases by 3.4\%–17.9\% across the two datasets. This confirms the necessity of each proposed component. Among them, the removal of dynamic graph construction (w Static) causes the most significant degradation, highlighting that dynamic graphs are the foundation for addressing microservice dynamism. Without dynamic alignment of temporal features, both TCD and SCO lose much of their effectiveness. Additionally, the performance degradation is generally more pronounced on D2 than on D1, which indicates that modeling temporal and dynamic characteristics becomes increasingly critical when the dataset involves more complex fault patterns and a larger number of services. Interestingly, the \textit{Avg@5} results on D1 show relatively minor decreases, suggesting that when the service scale is small, root causes are easier to locate within the top-5 ranking compared to the D2 dataset.

\begin{figure}[h]
  \centering
  \includegraphics[width=0.48\textwidth]{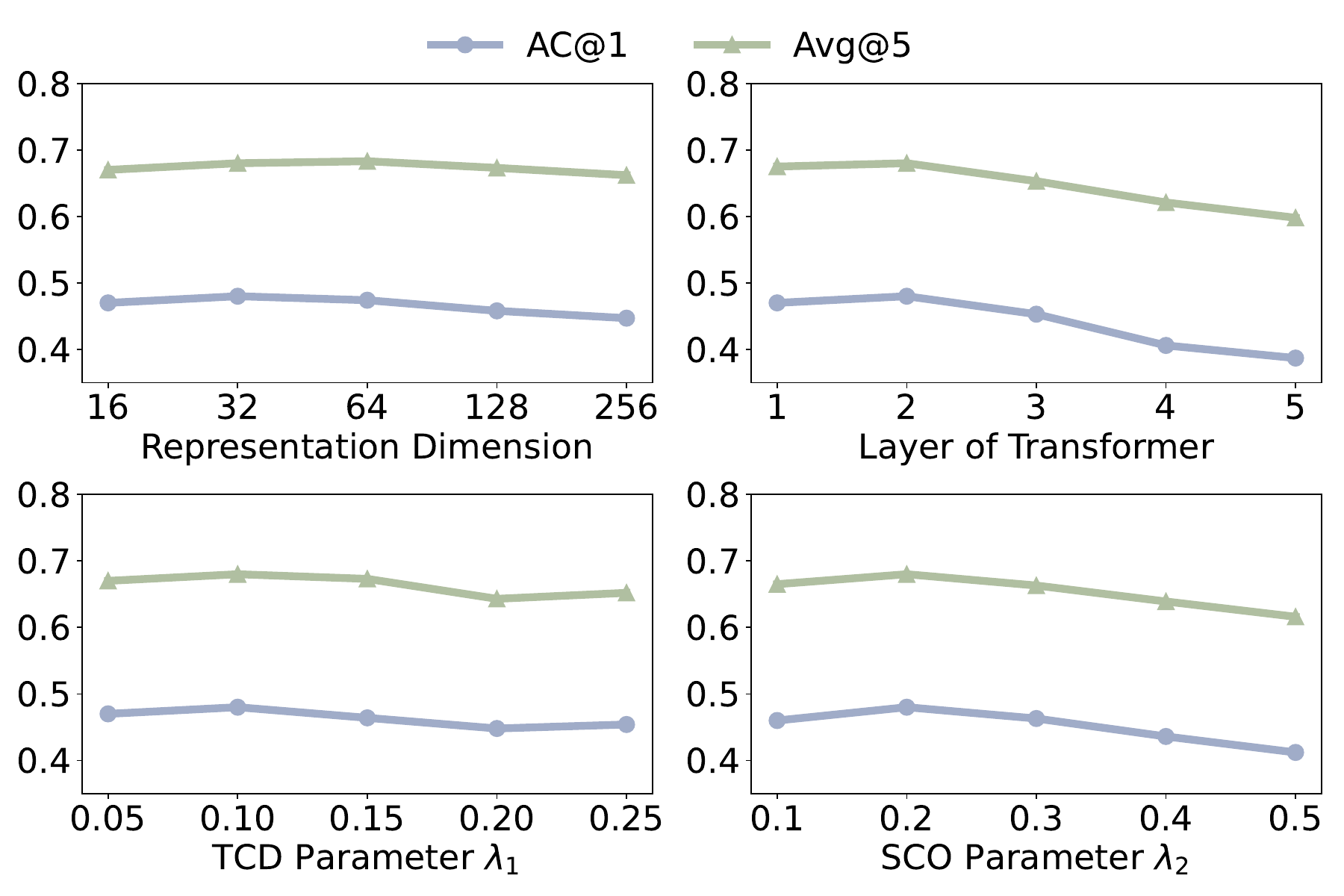}
  \caption{Parameter Sensitivity Analysis of \model\ on D2.}
  \label{fig4:parameter}
\end{figure}
\subsection{RQ3: Parameter Sensitivity}
We conducted a parameter sensitivity analysis of \model\ on the D2 dataset to examine the robustness of our method with respect to several key hyperparameters, including the representation dimension, the number of Transformer layers, the TCD parameter $\lambda_1$, and the SCO parameter $\lambda_2$. As shown in Figure~\ref{fig4:parameter}, each sub-figure reports the variations of \textit{AC@1} and \textit{Avg@5} as a function of these hyperparameters. 

\textbf{Overall, the results demonstrate that \model\ maintains stable performance within a reasonable range of parameter settings.} For the representation dimension, the model achieves its best performance when the dimension is set to 32, while the overall trend remains relatively flat, indicating insensitivity to larger or smaller values. A similarly stable trend can be observed for $\lambda_1$, where the results fluctuate only slightly across different values. In contrast, the number of Transformer layers has a more pronounced effect: performance peaks at 2 layers but decreases consistently when the depth exceeds this threshold. This degradation can be attributed to overfitting introduced by excessive Transformers stacking, which persists even with early stopping. Regarding the SCO parameter $\lambda_2$, the model performs best at $\lambda_2=0.2$; increasing $\lambda_2$ beyond this point leads to a noticeable decline in both \textit{AC@1} and \textit{Avg@5}. In summary, these findings suggest that \model\ is robust to a wide range of hyperparameter choices, but overly large parameter values, particularly in the number of Transformer layers and $\lambda_2$, may adversely affect its performance.

\section{Related Work}
In this section, we review related work following a progression from single-modality to multi-modality RCA methods.

Single-modality method using metrics, traces or logs for diagnosis. Metric-based approaches, such as GrayScope \cite{zhang2024illuminating}, infer service states by mining causal dependencies among metrics to localize root causes. Trace-based methods, exemplified by MicroRank \cite{yu2021microrank}, exploit distributed traces to construct call graphs and rank suspicious spans or services. Log-based techniques \cite{du2017deeplog, kobayashi2017mining} model message templates or log-event sequences to detect anomalous behavior. While these approaches have shown effectiveness in specific scenarios, they often fail to capture the holistic view required to disambiguate cascading failures in complex service dependencies.

Motivated by the limitations of single-modality analysis, recent studies have increasingly turned to multi-modality monitoring data, which provide a more comprehensive view of microservice health. Representative efforts include Eadro \cite{lee2023eadro}, which integrates logs, metrics, and traces to strengthen dependency modeling; ART \cite{sun2024art} and DiagFusion \cite{zhang2023robust}, which dynamically construct service graphs and apply graph neural networks to model spatial information; and M-RCA \cite{wang2024mrca}, which enhances fault localization by combining causal reasoning with reinforcement learning. These studies consistently demonstrate that integrating heterogeneous data sources substantially improves diagnostic accuracy and coverage compared to single-modality methods.

Despite these advances, most existing RCA techniques still assume static system structures and stationary data distributions, thereby limiting their applicability in real-world dynamic environments. In contrast, our work explicitly addresses the dynamic nature of microservices by introducing dynamic feature alignment, contrastive discrepancy learning, and root-cause-prioritized ranking, leading to more accurate and robust root cause analysis under dynamic conditions.
\section{Conclusion}
In this work, we propose \model, a dynamic-aware framework for root cause analysis in microservice systems.
By jointly modeling temporal dynamics and system-level interactions, \model\ effectively addresses
the cascading fault propagation in dynamic microservice environment.
Through a causal representation discrimination module that unifies contrastive and ranking mechanisms, \model\ achieves robust and interpretable fault localization against normal fluctuation confusion, and deviation–cause misalignment.
Comprehensive experiments on two benchmark datasets show that \model\ consistently surpasses all baselines, achieving an average \textit{AC@1} of 0.63 and absolute improvements of 0.25–0.46, while delivering accurate and interpretable diagnoses in dynamic microservice systems.

\end{document}